\documentstyle[twoside,fleqn,espcrc2]{article}

\newcommand{\BE}{\begin{eqnarray}}
\newcommand{\EE}{\end{eqnarray}}
\newcommand{\BDM}{\begin{displaymath}}
\newcommand{\EDM}{\end{displaymath}}
\newcommand{\NN}{\nonumber\\}

\newcommand{\spec}{\mbox{spec}}

\newcommand{\C}{C^\gamma}

\newcommand{\g}{\gamma}

\newcommand{\Ch}{Chebyshev }
\newcommand{\G}{Gegenbauer }
\renewcommand{\L}{Legendre }

\title{Fractional Inversion in Krylov Space}

\author{B. Bunk
      \address{Institut f{\"u}r Physik, Humboldt--Universit{\"a}t zu Berlin,
            Invalidenstr.110, D--10115 Berlin, Germany}
      \thanks{email: bunk@physik.hu-berlin.de}
      }

\begin{document}

\begin{abstract}
The fractional inverse $M^{-\gamma}$ (real $\gamma >0$) of a matrix $M$ is
expanded in a series of Gegenbauer polynomials. If the spectrum of $M$ is
confined to an ellipse not including the origin, convergence is exponential,
with the same rate as for Chebyshev inversion. The approximants can be
improved recursively and lead to an iterative solver for $M^\gamma x = b$ in
Krylov space. In case of $\gamma = 1/2$, the expansion is in terms of Legendre
polynomials, and rigorous bounds for the truncation error are derived.
\end{abstract}

\maketitle

\section{\G Polynomials}

The key relation is the generating function
\BE
      (1 + t^2 - 2 t z)^{-\g} = \sum_{n=0}^\infty t^n \C_n(z)
                                    \label{genfctC}
\EE
with a real parameter $\g > 0$. It defines the \G polynomials $\C_n(z)$,
see \cite{HTFII}, part 10.9. They obey the recursion relation
\BE
      (n+1) \C_{n+1}(z) + (n+2\g-1) \C_{n-1}(z) \NN
      = 2(n+\g) \, z\C_n(z)
                                    \label{Cn-recur}
\EE
($n \ge 0$, $\C_{-1} = 0$) and are normalised such that
\BE
      \C_n(1) = \frac{\Gamma(2\g + n)}{n! \, \Gamma(2\g)} .
\EE
Special cases are the \L polynomials
\BE
      C^{1/2}_n(z) = P_n(z)
\EE
and the \Ch polynomials of the second kind
\BE
      C^1_n(z) = U_n(z) .
\EE
Estimates of $\C_n$ are obtained with the aid of the integral representation
\BE
   \lefteqn{      \C_n(z) = \frac{2^{1-2\g} \Gamma(2\g+n)}
                                 {n! \, \Gamma(\g)^2}       }\NN
            &&    \int_0^\pi d\varphi \, (\sin\varphi)^{2\g-1}
                  (z + \sqrt{z^2 - 1} \cos\varphi)^n .
\EE

\section{Convergence of \G expansions}

The relation (\ref{genfctC}) was introduced to define the \G polynomials,
but it also serves as an efficient expansion
of $(1 + t^2 - 2tz)^{-\g}$ in polynomials of $z$. In this context,
the relative error of the truncated series is of particular importance: 
\BE
      R_n(z) &=& 1 -  (1 + t^2 - 2 t z)^\g \sum_{k=0}^n t^k \C_k(z)     \\
             &=& (1 + t^2 - 2 t z)^\g \sum_{k=n+1}^\infty t^k \C_k(z)
                        \label{Rn}
\EE

On the line segment $z \in [-1,1]$, the \G polynomials are uniformly
bounded by
\BE
      |\C_n(z)| &\leq& \C_n(1)  \quad\mbox{for}\quad z \in [-1,1]   \NN
            &\simeq& \frac{n^{2\g-1}}{\Gamma(2\g)}
                  \quad\mbox{as}\quad n\to\infty      \label{Cn-asympt}
\EE
Therefore the expansion (\ref{genfctC}) converges uniformly in $z$, provided
that $|t| < 1$. The relative error (\ref{Rn}) decreases like
\BE
      |R_n(z)| = {\cal O}(|t|^{n+1}) ,
                  \label{orderR}
\EE
possibly up to powers of $n$.
For the \L case ($\g = 1/2$), the following rigorous estimate is proven
in \cite{BB97}:
\BE
      |R_n(z)| \le |t|^{n+1} \quad\mbox{for}\quad z \in [-1,1] .
                  \label{boundR}
\EE

To extend these considerations to $z \not\in [-1,1]$, it is convenient to
parametrise the complex plane in terms of confocal ellipses (as in the case
of \Ch polynomials, see \cite{Mant77,BdF95}). Inside the ellipse
\BE
      z = \cosh(\theta + i\phi), \quad \theta \ge 0 ,\quad \phi \in [0, 2\pi]
                              \label{ellipse}
\EE
$\C_n$ is bounded by
\BE
      |\C_n(z)| &\le& \C_n(\cosh\theta)         \NN
            &\le& \C_n(1) \cosh n\theta
\EE
and convergence of the expansion (\ref{genfctC}) is uniform in $z$
if $|t| e^\theta < 1$. We also have
\BE
      |R_n(z)| = {\cal O}\left( (|t| e^\theta)^{n+1} \right)
\EE
and for $\g = 1/2$ the general bound\cite{BB97}
\BE
      |R_n(z)| \le (|t| e^\theta)^{n+1}.
                  \label{boundRc}
\EE

\section{Iterative solver}

The aim is to solve
\BE
      M^\g x = b        \label{Mx}
\EE
approximately in Krylov space, i.e. in terms of polynomials of $M$ acting
on $b$. To this end, parametrise $M$ as
\BE
      M = c (1 + t^2 - 2tA)   \label{M}
\EE
with $c,t \in C$. If $M$ has a spectrum on a line away from the origin,
we can find a transformation with $|t| < 1$ such that
$\spec A \subset [-1, 1]$. More generally, if the spectrum of $M$ is bounded
by an ellipse which does not include the origin, it can be mapped into an
ellipse (\ref{ellipse}) with $|t| e^\theta < 1$. In any case, $|t|$ resp.
$|t| e^\theta$ should be as small as possible for optimal convergence.

The formal solution
\BE
      x = c^{-\g} (1 + t^2 - 2tA)^{-\g} \, b
\EE
suggests to insert the \G expansion (\ref{genfctC}) and to form the
approximants
\BE
      x_n = c^{-\g} \sum_{k=0}^n t^k \C_k(A) \, b
          = \sum_{k=0}^n t^k s_k .
                        \label{xn}
\EE
The shifts
\BE
      s_n \equiv c^{-\g} \C_n(A) \, b
\EE
inherit the recursion relation (\ref{Cn-recur}):
\BE
      (n+1) s_{n+1} + (n+2\g-1) s_{n-1}   \NN
            = 2(n+\g) \, A s_n            \label{sn-recur}
\EE
($n \ge 0$) to be startet from
\BE
      s_{-1} &=& 0            \NN
      s_0 &=& c^{-\g} b .     \nonumber
\EE
As to the stability of the recursion, the error $\delta s_n$ evolves
for large $n$ according to the approximate relation
\BE
      \delta s_{n+1} + \delta s_{n-1} = 2 A \, \delta s_n .       \nonumber
\EE
Consider an eigenvalue $\lambda$ of $A$ and express it as
$\lambda = \cosh(\vartheta + i\varphi)$ with $0 \le \vartheta \le \theta$.
Then the error of the corresponding mode of $s_n$ behaves like
\BE
      \delta s_n \propto e^{\pm(\vartheta + i\varphi)n} .         \nonumber
\EE
Note that the shifts $s_n$ are to be multiplied by $t^n$ when accumulated
for the solution (\ref{xn}). Therefore, the modes of iterated errors in
$t^n \delta s_n$ contribute $\sim (|t|e^{\pm \vartheta})^n \le (|t|e^\theta)^n$,
i.e. error propagation is damped due to $|t|e^\theta < 1$.

In conclusion, the recursion (\ref{sn-recur}) provides a stable solver
(\ref{xn}) for the linear problem (\ref{Mx}).
This was confirmed in a first application\cite{Elser97} within the boson
algorithm for dynamical fermions.

A peculiar feature of this solver is that it cannot be started from an
arbitrary vector $x$, but only from $x_{-1} = 0$ or $x_0 = s_0$ as stated above.
This is due
to the fact that $M^\g x_{-1}$ cannot be evaluated for arbitrary $x_{-1}$
(which is needed to bring the start vector to the r.h.s. of the equation).
If an approximant $x_n$ is to be improved in further iterations, the last
shifts $s_{n-1}, s_n$ have to be saved as well.

\section{Error estimates}

The rest vector and the relative error of $x_n$ are both controlled
by $R_n$ eq.(\ref{Rn}):
\BE
      r_n &=& b - M^\g x_n = R_n(A) \, b    \\
      x - x_n &=& M^{-\g} r_n = R_n(A) \, x
\EE
Note that, in contrast to the case of plain inversion ($\g = 1$), $r_n$ is not
available during the iterative process, because $M^\g$ is not computable
for fractional $\g$. Therefore mathematical bounds for $\|R_n(A)\|$ are
important to estimate the quality of the approximation. (Throughout this
paper, $\|\cdot\|$ designates the Euclidean norm for vectors and the induced
matrix norm.)

If $M$ is normal, i.e. $[M,M^\dagger] = 0$, the same is true for $A$, and
$\|R_n(A)\|$ is determined by the spectrum of $A$. This results in an
estimate for the relative error
\BE
      \|x - x_n\| / \|x\| &\le& \|R_{n+1}(A)\|        \NN
      &=& \max_i |R_n(\lambda_i(A))| .
\EE
At this point, uniform bounds like (\ref{boundR}) or (\ref{boundRc}) over the
spectrum of $A$ are useful. So far, they are available for $\g = 1/2$ only.

As an example, let $M = M^\dagger$ be positive definite with
$\spec M \subset [\lambda_{min}, \lambda_{max}]$. $c$ and $t$
are chosen such that eq.(\ref{M}) maps $M = \lambda_{max} \to A=-1$ and
$M = \lambda_{min} \to A=1$. This implies
\BE
      \kappa &\equiv& \frac{\lambda_{max}}{\lambda_{min}}
            = \left( \frac{1+t}{1-t} \right)^2        \NN
   \Rightarrow
      t &=& \frac{\sqrt{\kappa} - 1}
                 {\sqrt{\kappa} + 1}                  \nonumber
\EE
Eq.(\ref{orderR}) shows that $t$ is the convergence factor, and it agrees
with the one for optimal
\Ch inversion and the bound for the Conjugate Gradient solver.

If $\spec M$ is not known, one may proceed as follows:
define $A$, based on a guess about the spectrum, and monitor the
norm of the shifts $\|s_n\|$. The estimate
\BE
      \|s_n\| / \|s_0\| &=& \|\C_n(A)b\| / \|b\|      \NN
            &\le& \|\C_n(A)\|                         \NN
            &\le& \C_n(\cosh\theta)
\EE
is assumed to be saturated for large $n$, and this allows for a
(pragmatic) determination of $\theta$. If $|t|e^\theta \ge 1$, there will be
no convergence and one has to stop and try again. Otherwise, $\theta$ is employed
in the error estimate.
It may also help to refine the parametrisation of $M$ (choice of $c$ and $t$)
for later use.

\section{Conclusions}

\begin{itemize}
\item
      The \G expansion allows to construct iterative solvers
      for $M^\g x = b$ with the same rate of convergence as
      (optimal) \Ch methods for $\g = 1$.
\item
      If the spectrum of $M$ is known, the relative error of the
      approximation can be estimated. Tight bounds are known for $\g = 1/2$.
\item
      If the spectrum is uncertain, the growth of the shift vectors can
      be used for a rough estimate.
\item
      Polynomial approximations ${\bf P}_n(M) \approx M^{-\g}$ are obtained
      along the same lines.
\item
      The polynomials generated by the \G expansion are not optimal in
      any sense, but the method is distinguished by its conceptional and
      computational simplicity.
\end{itemize}


\begin{thebibliography}{9}

\bibitem{HTFII}   A. Erd\'elyi ed., Higher Transcendental Functions,
                        Vol. II, McGraw-Hill, New York 1953.
\bibitem{BB97}    B. Bunk, to be published.
\bibitem{Mant77}  T. A. Manteuffel, Numer. Math. 28 (1977) 307.
\bibitem{BdF95}   A. Bori\c{c}i, Ph. de Forcrand, Nucl. Phys. B454 (1995) 645.
\bibitem{Elser97} S. Elser and B. Bunk, Nucl. Phys. B (Proc. Suppl.) 63
                        (1998) 940.

\end{thebibliography}
\end{document}